# Accelerating Materials Development *via* Automation, Machine Learning, and High-Performance Computing

Juan Pablo Correa-Baena[1], Kedar Hippalgaonkar[2], Jeroen van Duren[3], Shaffiq Jaffer[4], Vijay R. Chandrasekhar[5], Vladan Stevanovic[6], Cyrus Wadia[7], Supratik Guha[8], Tonio Buonassisi[1*]

[1]Massachusetts Institute of Technology, Cambridge, MA 02139, USA

[2]Institute of Materials Research and Engineering (IMRE), A*STAR (Agency for Science, Technology and Research), Innovis, Singapore

[3]Intermolecular Inc., San Jose, CA 95134, USA

[4]TOTAL American Services, Inc., 82 South Street, Hopkington, MA 01748, USA

[5]Institute for Infocomm Research (I$^2$R), A*STAR (Agency for Science, Technology and Research), #21-01 Connexis (South Tower), Singapore

[6]Colorado School of Mines, Golden, CO 80401, USA

[7]Lawrence Berkeley National Laboratory, Berkeley, CA 94720, USA

[8]Center for Nanoscale Materials, Argonne National Laboratory, Argonne, IL 60439, USA

*Corresponding author: Tonio Buonassisi, buonassisi@mac.com

**Successful materials innovations can transform society. However, materials research often involves long timelines and low success probabilities, dissuading investors who have expectations of shorter times from bench to business. A combination of emergent technologies could accelerate the pace of novel materials development by 10x or more, aligning the timelines of stakeholders (investors and researchers), markets, and the environment, while increasing return-on-investment. First, tool automation enables rapid experimental testing of candidate materials. Second, high-throughput computing (HPC) concentrates experimental bandwidth on promising compounds by predicting and inferring**

**bulk, interface, and defect-related properties. Third, machine learning connects the former two, where experimental outputs automatically refine theory and help define next experiments. We describe state-of-the-art attempts to realize this vision and identify resource gaps. We posit that over the coming decade, this combination of tools will transform the way we perform materials research. There are considerable first-mover advantages at stake, especially for grand challenges in energy and related fields, including computing, healthcare, urbanization, water, food, and the environment.**

The development of novel materials has long been stymied by a mismatch of time constants (**Figure 1**). Materials development typically occurs over a 15–25-year time horizon, sometimes requiring synthesis and characterization of millions of samples. However, corporate and government funders desire tangible results within the residency time of their leadership, typically 2–5 years. The residency time for postdocs and students in a research laboratory is usually 2–5 years; when a project outlasts the residency of a single individual, seamless continuity of motivation and intellectual property is often the exception, not the rule. Market drivers of novel materials development, informed by business competition and environmental considerations, often demand solutions within a shorter time horizon. This mismatch in time constants results in a historically poor return-on-investment of energy-materials (cleantech) research relative to comparable investments in medical or software development.[1]

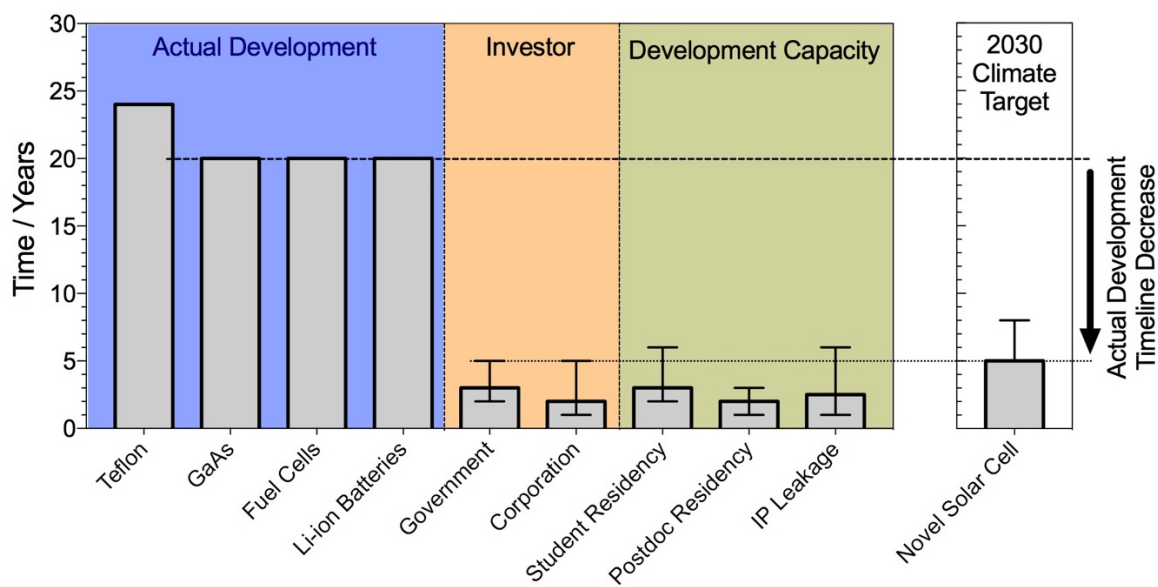

**Figure 1. Timelines for materials discovery and development.** Timelines of examples of certain technologies (blue area), typical academic funding grants (orange), development capacity (green) and deployment of sustainable energy (*i.e., via* solar cells) to fulfill the 2030 climate targets.

To bridge this mismatch in time horizons and increase the success rate of materials research, both public- and private-sector actors endeavor to develop new paradigms for materials development. The U.S. Materials Genome Initiative focused on three "missing links": computational tools to focus experimental efforts in the most promising directions, data repositories to aggregate learnings and identify trends, and higher-throughput experimental tools.[2] This call to action was mirrored in industry and by university- and laboratory-led consortia, many focused on simulation-based inverse design and discovery and properties databases. As these tools matured, the throughput of materials prediction often vastly outstripped experimentalists' ability to screen for materials with low rates of false negatives.

Today, a new paradigm is emerging for experimental materials research, which promises to enable more rapid discovery of novel materials.[3,4] **Figure 2** illustrates one such prototypical vision, entitled "accelerated materials development and manufacturing." Rapid, automated feedback loops are guided by machine learning, and an emphasis on value creation through end-product and industry transfer. There is a unique opportunity today to develop these capabilities in testbed fashion, with considerable improvements in research productivity and first-mover advantages at stake.

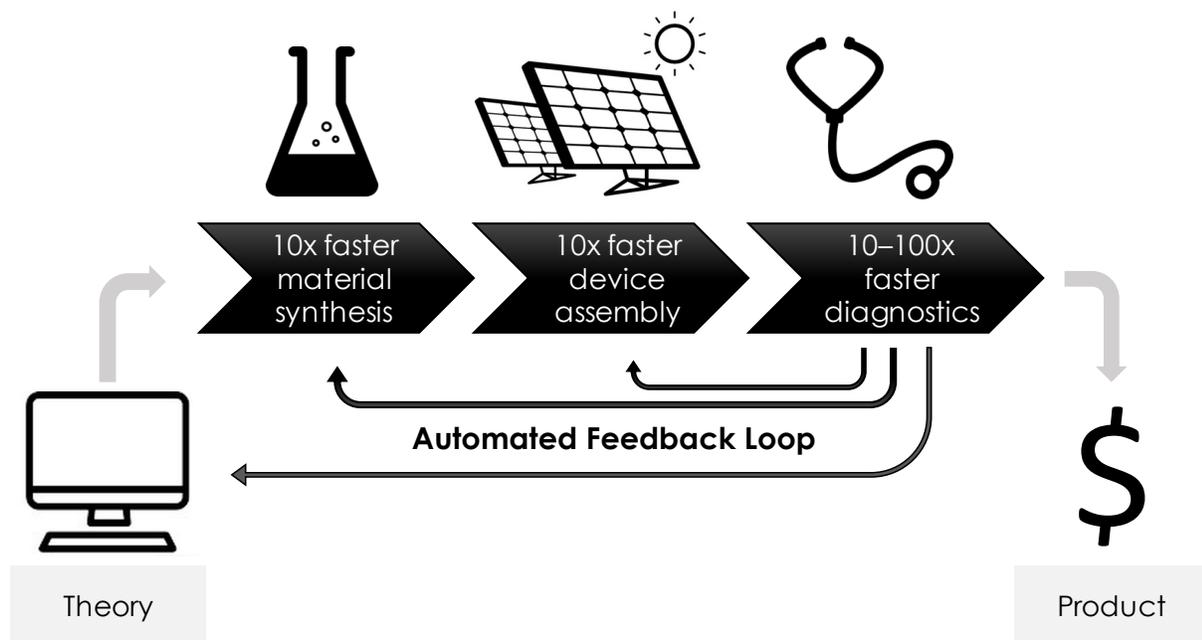

**Figure 2.** Schematic of accelerated materials discovery process. The automated feedback loop, driven by machine learning, drives process improvement. The theory, synthesis, and device processes take advantage of high-performance computing and materials databases. Icons from Ref. 31.

As is often the case with convergent technologies, one observes significant advances in individual "silos" before the leveraged ensemble effect bears its full impact. A historical example is three-dimensional printing, wherein 3D computer-aided design (CAD), computer-to-hardware interface protocols, and ink-jet printing technologies evolved individually, before being combined by Prof. Ely Sachs and his MIT team into the first 3D printer. The ability to observe emergent technologies within individual silos, and assemble them into an ensemble that is greater than the sum of its parts, mirrors the challenge in novel materials development today. The following paragraphs describe the discrete, emergent innovations in "siloed" domains that are presently converging, and promise to enable this paradigm shift within the next decade.

**Theory**: Today, the rate of theoretical prediction vastly outstrips the rate of experimental synthesis, characterization, and validation.[5] This emergence is enabled by three trends: faster computation, more efficient and accurate theoretical approaches and simulation tools, and the ability to screen large databases quickly, such as MaterialsProject.org. To bridge the growing gap between theory and experiment, researchers are increasingly focusing efforts on predictive materials synthesis routes, especially synthesis routes that consider environmental factors (*e.g.*, humidity), reaction-energy barriers, and kinetic limitations (so-called "non-equilibrium" synthesis).[17] In parallel, theorists seek to rationally design materials with combinations of properties — first, by predicting combinations of properties (*e.g.*, chemical, microstructural, interface, surface…) in one simulation framework and/or database, then connecting material predictions with device performance & reliability predictions, then extending this framework to both known and not-yet-discovered compounds, and ultimately, solving the inverse problem.

**High-Throughput Materials, Device, and Systems Synthesis**: Historically, slow vacuum-based deposition methods inhibit materials development. Modern vacuum-based tools, including combinatorial approaches and large-scale, fast serial deposition/reactions, enable meaningful rate increases for materials and device synthesis.[29,30] Variants of existing deposition methods (*e.g.*, close-space sublimation) offer higher growth rates, point-defect control, and precise stoichiometry and impurity control for process-compatible materials. Solution synthesis has gained acceptance with the emergence of higher-quality precursors and materials, including CdS quantum dots, polymer solar cells, and lead-halide perovskites.[5,6] The growing diversity of precursors (from molecular to nanoparticle), synthesis control (including solvent engineering), and thin-film synthesis methods (lab-based spin-coating to industrially-compatible large-area printing) makes this a powerful and flexible platform to deposit a range of new materials. Emergence of 3D printed materials provides another ubiquitous alternative. At laboratory scale, throughputs for such rapid synthesis routes[5,7] can be up to an order of magnitude greater than vacuum-based techniques, and remain to be explored for multinary materials with novel microstructures. With declining component costs and greater adoption of standards, the ability to rapidly combine discrete devices into components and systems in a modular and flexible manner is emerging.

**Defect Tolerance & Engineering**: Often, theoretical predictions are made for "ideal" materials systems. However, real samples contain defects (impurities, structural defects…), which can harm (or, occasionally, benefit) bulk and interface properties. To mitigate the risk of defect-induced false negatives during high-throughput materials screening, it is desirable to identify classes of materials less adversely affected by defects (so-called "defect tolerant" [8,9]), and rapidly diagnose & decouple the effects of defects on material performance. A notable recent example is the serendipitous discovery of lead-halide perovskites for optoelectronic applications.[6,7] In addition to being amenable to high-throughput solution-phase deposition, lead-halide perovskites also required orders of magnitude *less research effort* to achieve similar performance improvements to traditional inorganic thin-film materials (Figure 3). It is suspected that part of the facility to improve performance is owed to increased defect tolerance of lead-halide perovskites, resulting in improved bulk-transport properties. Determining the underlying physics of and developing design rules for defect tolerance may inform screening criteria for new materials, especially with new computational tools such as General Adversarial Networks (GANs) that are state-of-the-art in anomaly detection.[22,23] The next step lies in focusing experimental effort on candidates capable of rapid performance improvements during early screening and development, and wider process tolerance in manufacturing. In relation to the beneficial aspects of defects and impurities, recent theory advancements[15] in combination with computational tools to rapidly assess and predict solubility and electrical properties of defects[16] allows high-throughput screening of materials for applications where the desired functionality is enabled by the defects and/or dopants (*e.g.*, thermoelectrics, transparent electronics…).

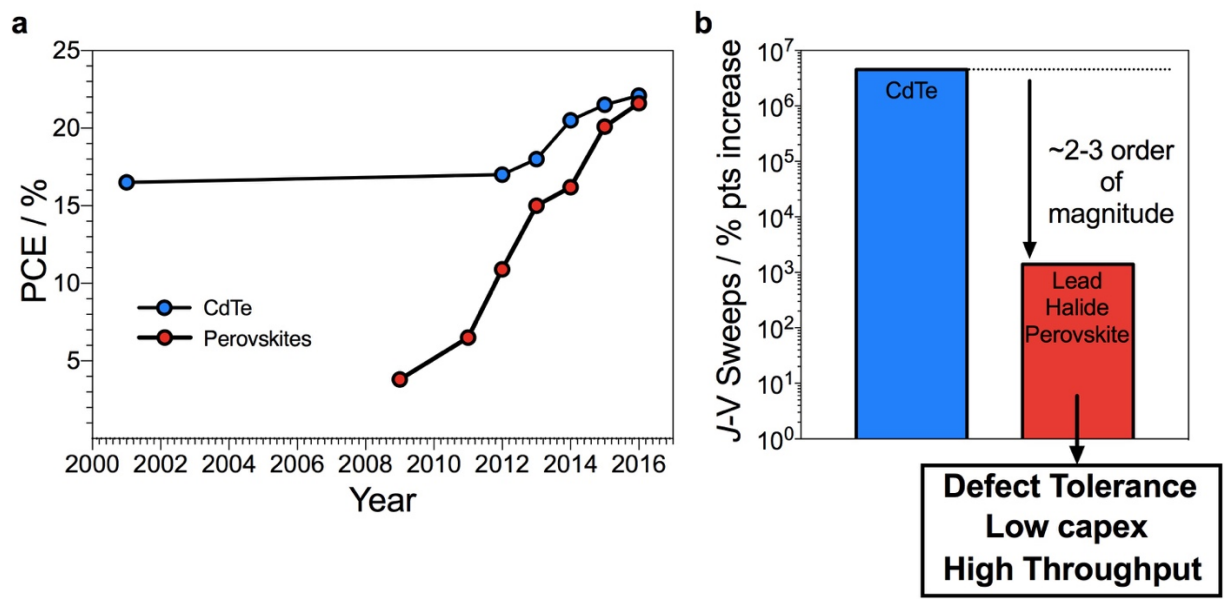

**Figure 3. A case study of fast materials development based on photovoltaic applications. a.** certified power conversion efficiency (PCE) over time for CdTe and perovskite solar cells. **b.** Number of *J*-V sweeps measured divided by the increase in percentage point achieved during the device development of CdTe and perovskite solar cells. Three orders of magnitude fewer *J-V* sweeps per percentage efficiency improvement were needed to advance perovskite efficiencies relative to traditional thin-film solar cell materials. We hypothesize that this difference is partially due to greater "defect tolerance" of perovskites, enabling a faster and more economical materials development process.

**High-Throughput Diagnosis**: Characterization tools have also benefitted from high-throughput computing, automation, and machine learning. For instance, one high-resolution X-ray photoelectron spectroscopy spectrum could take an entire day with technology from the 1970's, while the same measurement today requires less than an hour. Today, advanced statistics and machine learning promises to further accelerate the rate of learning. Tools now exist that can acquire multiple XPS spectra on a single sample (*e.g.*, with composition gradients), and automated spectral analysis of large datasets is now possible, enabling estimation of unknown materials in a compositional map. Others seek to replace spectroscopy with rapid non-destructive testing; several bulk and interface properties can be simultaneously diagnosed by using Bayesian inference in combination with non-destructive device testing, enabling ≥10x faster (and in certain cases, more precise) diagnosis *vis a vis* traditional characterization tools.[10] This kind of parameter estimation can be applied to finished components, devices, and systems, and has the potential to not only enable faster troubleshooting, but also to accurately estimate ultimate performance potential, thus informing the decision to pursue or abandon further investment in a given candidate material even at early stages of materials screening.

**Machine Learning** comprises a broad class of approaches, which may play several different roles in the future materials-development cycle. First, a common application of machine learning is for materials selection, in which historical experimental observations are used to inform predictions of future properties (attributes) of unknown compounds, or discover new ones.[24] Such an approach has been realized to help discover novel active layers in organic solar cells[11] and light-emitting diodes[12], and metal alloys[13,27], among many others.[28] Second, machine learning tools can help extract greater and more accurate information from diagnosis, as detailed in the previous section. Third, machine learning tools may help close the automation loop between *diagnosis* and *synthesis*, shown in **Figure 2**, by reducing the degree of human intervention and reliance on heuristics. For example, when relationships between experimental inputs and diagnosis outputs can be inferred by neural networks, detailed process and device models may no longer be needed to predict outcomes and optimize processes. All three applications of machine learning to the materials development cycle benefit from the availability of more data, to train and sharpen the predictive capacity of such tools.

Achieving predictability without losing physical insights is an emergent challenge and research opportunity. Such methods may also increase learning from diagnosis, by consolidating research output in singular databases, drawing automated inferences from the data, and in the future perhaps aggregating the experience and knowledge base *via* natural language processing of existing research papers and materials property databases.

**Envisioning the "Hardware Cloud":** Materials synthesis equipment today is becoming increasingly remotely operable—enabling research and operation by an investigator who is not in proximal presence to the deposition equipment. This opens up two related opportunities with far-reaching consequences. Large, expensive, synthesis equipment can be grouped together with massively parallel characterization equipment to form synthesis centers of the future, which are operated by remote users and researchers and managed by an on-site professional staff. Akin in concept to the Software Cloud concept, where one's computing and data is stored across machines worldwide in a seamless manner, a Hardware Cloud would enable a user to deposit, measure and carry out research (with real time feedback through in-situ characterization tools) across a number of networked materials processing systems distributed nationally or internationally in a seamless manner. This also leads to the second opportunity: to be able to store, curate, access, process and diagnose all data gathered in these networked experiments in Public or Private Clouds. (Protocols and formats for such science data collectives will be discussed in the following paragraphs.) This will greatly facilitate two emerging issues: (a) increasing the efficient availability of data across a wide number of experiments and experimental platforms for post-analysis; and (b) making available for analysis data that indicates "what did not work" — this is not easily available but is instrumental in the learning process, and has its own value in increasing the collective efficiency of research progress.

**Infrastructure Investments Toward Accelerated Materials Development and Manufacturing:** Realizing the vision shown in **Figure 2** requires a sustained commitment over several years to develop software, hardware, and human resources, and to connect these new capabilities in testbed fashion.

*Investments in applied machine learning*: Supported by ample investments into machine-learning methods development, a pressing challenge is how to down-select and apply the most appropriate machine-learning methods to enable the "automated feedback loop" shown in **Figure 2**. Compared to other widely recognized applications of machine learning today (*e.g.*, vision recognition, natural-language processing, and board gaming), materials research often involves sparse data sets (*e.g.*, small sample sizes and number of experimental inputs & outputs, for training and fitting) and less well-constrained "rules" (*e.g.*, complex physics and chemistry, non-binary inputs and outputs, large experimental errors, uncontrolled input variables, and incomplete characterization of outputs, to name a few). These realities make the typical materials-science problem (*e.g.*, layer-by-layer atomic assembly of a thin film) decidedly more complex and less well defined than a match of "Go," where the rules and playing board are constrained. Deep machine learning (DML) appears well poised to address this complexity. Computation speed can be improved by developing

"pre-trained" neural networks that incorporate the underlying physics and chemistry common to materials synthesis, performance, and defects, bringing DML within reach of commonly available hardware and software.

A balance must be found between achieving actionable results and inferring physical insight from "black-box" computational methods, to advance both engineering and scientific objectives, and minimize unintended consequences. There is a need to apply "white box" (*i.e.*, opposite of black box) machine learning methods to materials science problems. One possible approach may be application of semi-supervised deep learning algorithms, which learn with lots of unlabeled data and very little labeled data.[25]

Lastly, the ability of machine-learning tools to adapt to uncontrolled and changing experimental conditions is essential. Promising developments include online deep learning, which builds neural networks on the fly, gradually adding neurons (*e.g.*, as baseline experimental conditions change, or as new physics becomes dominant).[26]

*Investment in standards governing data formatting and storage* would facilitate data entry into machine-learning software. Standards embed contextual know-how, hierarchy, rational thought. Some communities have implemented standards governing raw and processed data, *e.g.*, crystallography, genetics, and geography. However, in most materials-research communities, there are no universally accepted and implemented data standards. Several materials databases have been created, often specialized by material class or application, and with varying protocols for updating information and enforcing hygiene. Furthermore, these databases often lack ability to quickly & accurately predict device-relevant *combinations* of properties (*e.g.*, chemical, mechanical, optoelectronic, microstructural, surface, interface…). Several data standards have been proposed[19–21]; widespread adoption may hinge on widespread adoption of data-management systems described in the next paragraph. In the absence of data standards, it is possible that the burden of data aggregation will shift onto natural language processors[18], *i.e.*, computer programs designed to extract relevant data from available media (*e.g.*, publications, reports, presentations, and theses).

*Investment in data-management tools* (*e.g.*, informatics systems) are needed to manage data obtained from lab equipment and store records, coordinate tasks, and enforce protocols. On one hand, such systems have been shown to be of high value for well-defined research problems and tool sets. For early-stage materials research, data management tools require a deft balance between flexibility and standardization, and the ability to accommodate non-standard workflows, multiple participants, and equipment spread across multiple sites, including shared-use facilities, in an elegant and seamless manner. When implemented well, data-management systems can increase the quality, uniformity, and accessibility of data serving as inputs into machine-learning tools; when implemented too inflexibly, data-management systems can cause frictions to researcher workflow and stimulate their resistance. It is possible that, as suggested by Rafael Jaramillo (MIT), metadata-based distributed data-management systems may warrant strong consideration for early-stage materials research; a challenge will be, how to capture metadata in an automated, accurate, thorough, and comprehensive manner.

*Investments in infrastructure* are needed, to increase throughput of synthesis, device-fabrication, and diagnosis tools. The potential of automation must be realized, without sacrificing material quality and offsetting the advantages of higher throughput with an increase in false negatives. The emergence of multi-parameter estimation methodologies, including Bayesian inference and Design of Experiments (DoE) algorithms, invites the invention new non-destructive diagnostic apparatus designed to take full advantage of these new methodologies.

There are significant challenges associated with producing and analyzing large quantities of data. New tools being developed by machine learning specialists invite the possibility of *modifying hardware design to take advantage of machine-learning tools*, rather than the other way around.

Revised policies at institution, funding agency, and government levels may accelerate or stymie the required ongoing investments at levels large and small, and invites considering how export control laws, import duties, grant purchasing restrictions, overhead rates, auditing, and claw-back clauses affect required equipment investments to enable this transformation.

**Human-Capital Investments Toward Accelerated Materials Development and Manufacturing:**

Investments in human capital are required to prepare researchers to leverage these new tools. The transition from being "data-poor" to being "data-rich" invites changes in how we *think*, how we *incentivize*, and how we *teach*.

*How we think*: In a "data-poor" world, the time and cost of conducting each experiment is relatively large, and a risk-adverse mindset is advantageous. In a "data-rich" world, a larger number of unique experiments can be conducted per unit time, meaning that failure of any given experiment will have lesser negative impact on a researcher's milestones and publication record. This will enable researchers to experiment with greater creativity and risk-taking. This has three important implications for "how we think": First, a greater premium will be placed on experimental concept and design, as researchers who design experiments amenable to new tools will be rewarded. Second, a decreasing cost-per-experiment may result in reduced barriers for junior researchers to establish themselves, decreasing the premium of initial investment, prompting new as well as established researchers to explore new fields.

Third, an accelerated materials development framework invites a system-level perspective[14] that mirrors the new tools. Greater experimental throughput suggests that devices and systems may increasingly be analyzed holistically in lieu of isolated sub-components, test structures, and proxies. A "data-rich" world will allows us to analyze complex systems more directly, with lesser need to break into sub-components or impose *a priori* simplifications even without complete visibility into each sub-component. Wielding these new computer-based tools to greatest effect requires that researchers learn to "think" like machine-learning algorithms, appreciating the nuances and trade-offs of different approaches, requiring a mindset change providing an opportunity to identify weak links faster, focusing effort on those parameters with highest returns on investment.

*Incentives*: Encouraging the mindset change and transitions mentioned in the previous section will be complemented with a "constant of friction" governed in part by professional incentives of decades-old institutions. Young researchers will be encouraged to take proactive steps if they are rewarded by hiring committees, promotion committees, fellowship & awards committees, journal editors, and conference committees. Funding agencies could encourage open-source development of equipment that enables integration of high throughput synthesis of materials with data management. Industries may see value in funding solution-driven system-level approaches to accelerate their development timelines.

*Community*: Realizing this future requires merging domain expertise currently resident in robotics, software, computer science, electronics, materials, and design silos, each with their own language / acronyms, and academic conferences. The learning curve to become even a generalist in these different domains remains very steep. Reducing barriers to communication and achieving percolation of ideas across domains may be facilitated *via* cross-cutting conferences, workshops, and creation of funded research centers. Adoption of best practices across various fields can be encouraged *via* these percolation pathways of ideas.

*Education and Up-Skilling*: Public opinion (read: support or opposition) to ML/AI is influenced by whether or not citizens can envision a hopeful future that includes their employment and empowers society. First, these transformations require individuals at all levels and employment types to be willing to up-skill. Educators at all levels have an opportunity to revamp their curricula, considering both technical and societal impacts. Online tools and courses for machine-learning / artificial intelligence are growing in availability, but direct applications to materials science and systems engineering are needed. Second, we are invited to consider *how we teach* reflects the most suitable skills and mindsets to harness the full potential of accelerated materials development & manufacturing platforms. Domain expertise in supporting fields, including advanced statistics, will increase in utility with the mainstreaming of system-level design of experiments. Third, the scientific method will still be valid, and the premium will only increase for asking the right questions, designing good experiments, and disseminating results well.

**Conclusions**

The convergence of high-performance computing, automation, and machine learning promises to accelerate the rate of materials discovery, better aligning investor and stakeholder timelines. These new tools are set to become an indispensable part of the scientific process. >10x faster synthesis, device fabrication, diagnostics in a (semi-)automated feedback loop are distinctly possible in the near future. Discrete advances in theory, high-throughput materials, device, systems synthesis, diagnostics, the understanding of defects and defect tolerance, and machine learning are enabling this transition. There are several infrastructure and human-capital needs to enable this future, including greater emphasis on appropriate applications of existing methods to materials-relevant problems, adoption of data and metadata standards, data-management tools, and laboratory infrastructure, including both decentralized and centralized facilities. To integrate these tools into the R&D ecosystems depends in part on several human elements — namely, the time needed to evolve incentive structures, community support, education & up-skilling offerings, and researcher mindsets, as our field transitions from thinking "data poor" to thinking "data rich." We envision a scientific laboratory where the process of materials discovery continues without disruptions, aided by computational power augmenting the human mind, and freeing the latter to perform research closer to the speed of imagination, addressing societal challenges in market-relevant timeframes.

**Acknowledgements**: The ideas represented herein evolved in discussion with numerous individuals, including but not limited to Riley Brandt, Danny Ren, Felipe Oviedo, Daniil Kitchaev, Rachel Kurchin, I. Marius Peters, Shijing Sun, Rafael Jaramillo, Gang Chen, and Anantha Chandrakasan of MIT/SMART; Benjamin Gaddy of Clean Energy Trust; Rolf Stangl, Chaobin

He, and Anthony Cheetham of NUS; Dirk Weiss and Raffi Garabedian of First Solar; BJ Stanbery of Siva Power; Karthik Kumar, Sir John O'Reilly, Pavitra Krishnaswamy, Alfred Huan, and Cedric Troadec of A*STAR; Andrij Zakutayev, Stephan Lany, Dave Ginley, Greg Wilson, and William Tumas of NREL; and Lydia Wong, Shuzhou Li, Tim White, and Subbu Venkatraman of NTU, among many others.